\documentclass[3p,twocolumn]{elsarticle}

\usepackage[]{graphicx}
\graphicspath{{fig/}}

\usepackage{placeins}

\usepackage{array}
\newcolumntype{C}[1]{>{\centering\arraybackslash}p{#1}} 

\newif\ifbibtex
\bibtextrue

\ifbibtex

\usepackage{natbib}
\else

\expandafter\let\csname ver@natbib.sty\endcsname\relax
\let\citeyear\relax
\let\citeauthor\relax
\let\citet\relax
\let\citep\relax

\makeatletter
\let\c@author\relax
\makeatother

\usepackage[backend=biber,natbib,style=authoryear,dashed=false,maxbibnames=3,minbibnames=3,maxcitenames=3,uniquelist=false,maxalphanames=1,labelalpha,sorting=anyt,doi=false,isbn=false,url=false,eprint=true]{biblatex}

\AtEveryBibitem{\clearfield{title}}  

\defbibenvironment{bibliography}{\enumerate{}{}}{\endenumerate}{\item}

\DeclareNameAlias{sortname}{last-first}

\usepackage{filecontents}

\DeclareSourcemap{ 
  \maps[datatype=bibtex]{
    \map{
      \step[fieldsource=adsurl, match={http://adsabs.harvard.edu/abs/},
        replace={}]
    }
  }      
}

\DeclareFieldFormat{adsurl}{Bibcode: \href{http://adsabs.harvard.edu/abs/#1}{#1}}

\renewbibmacro{finentry}{\printfield{adsurl}\finentry}

\fi

\usepackage{aas_macros}

\usepackage{ifpdf}
\ifpdf
  \usepackage[pdftex]{hyperref}
  \hypersetup{
    colorlinks,%
    citecolor=black,%
    filecolor=black,%
    linkcolor=black,%
    urlcolor=cyan}
\else
  
  \newcommand\url[1]{\texttt{#1}}
\fi

\begin{document}

\title{The LOFAR Solar Imaging Pipeline and the LOFAR Solar Data Center}

\author{F.~Breitling}
\author{G.~Mann}
\author{C.~Vocks}
\author{M.~Steinmetz}
\author{K.~G. Strassmeier}

\address{Leibniz-Institut f\"ur Astrophysik Potsdam (AIP), An der Sternwarte 16, 14482 Potsdam, Germany}

\renewcommand*{\today}{September 28, 2015}

\begin{abstract}
LOFAR is a new and sensitive radio interferometer that can be used for dynamic
high-resolution imaging spectroscopy at low radio frequencies from 10 to 90
and 110 to 250 MHz. Here we describe its usage for observations of the Sun and
in particular of solar radio bursts. We also describe the processing,
archiving and accessing of solar LOFAR data, which is accomplished via the
LOFAR Solar Imaging Pipeline and the LOFAR Solar Data Center.
\end{abstract}


\maketitle

\section{Introduction}
The Sun is a low to moderate activity star and because of its proximity to
Earth a unique astrophysical object for studying stellar phenomena in great
detail. Radio observations can make substantial contributions to such studies
since the Sun is an intense radio source. Its $10^6$ K hot corona emits
thermal radio radiation. In addition, solar activity produces intense
non-thermal radio radiation observed as radio bursts
\citep{1985srph.book...37M,2005LNP...656...49W,2006GMS...165..221M}. They are
caused by a release of energy during a reconfiguration of the Sun's magnetic
field which also results in flares and eruptive events such as coronal mass
ejections. The radiation is generated by plasma emission and emitted near the
local plasma frequency and/or its harmonics \citep{1985srph.book..177M}. For
instance, type III bursts are a typical phenomenon in solar radio radiation
and are signatures of energetic electrons propagating along magnetic field
lines in the corona
\citep{1950AuSRA...3..541W,1985srph.book..289S,2011pre7.conf..373B}. Since the
radio emission occurs near the local electron plasma frequency $f_\mathrm{pe}
= (e^{2}N_\mathrm{e}/\epsilon_{0}m_\mathrm{e})^{1/2}/2\pi$ ($e$, elementary
charge; $\epsilon_{0}$, permittivity of free space; $m_\mathrm{e}$, electron
mass), and because of the gravitational density stratification of the corona,
higher and lower frequencies are emitted in the lower and higher corona,
respectively.

With the Low Frequency Array (LOFAR, \cite{2013A&A...556A...2V}) a sensitive
high resolution radio interferometer became available for radio observations
in the frequency range from 10 to 250~MHz corresponding to a radial distance
between 1 and 3 solar radii ($R_\odot$) in the corona
\citep{1999A&A...348..614M}. Its capability for high resolution dynamic
imaging spectroscopy makes it particularly useful for spatial and time
resolved observations of solar radio burst. The LOFAR Key Science Project
``Solar Physics and Space Weather with LOFAR'' (Solar KSP,
\cite{2011pre7.conf..507M}) was formed to use LOFAR for solar observations and
to address open questions. Its goals are the coordination of solar
observations, the solar data processing and analysis and the provision of the
resulting data products to the scientific community. These goals led to the
development of the LOFAR Solar Imaging Pipeline and the LOFAR Solar Data
Center described here.

\section{LOFAR}
LOFAR \citep{2013A&A...556A...2V} is a European digital radio interferometer
developed under the leadership of the Netherlands Institute for Radio
Astronomy (ASTRON). It currently consists of 48 antenna stations distributed
over the Netherlands (40), Germany (5), United Kingdom (1), France (1) and
Sweden (1) with its center near Netherlands city Groningen. Additional
stations are currently under constructions in Poland (3) and Germany (1) or in
the planning phase. A LOFAR station consists of high- and low-band antenna
fields and an electronics container with the receiver and computer hardware.
An antenna field is composed of identical antennas, which are added to a
station beam (or pointing) to receive the signal from a specific sky
direction. The data of the station pointings are sent to the correlator in
Groningen which forms the array or sub-array pointings.

LOFAR operates in the low frequency range from $10-90$ (low-band) and
$110-250$~MHz (high-band) in full polarization with high sensitivity and
resolution for imaging spectroscopy. LOFAR's beam forming allows for the
simultaneous observation of several different sky regions, which is important
for the calibration of highly variable radio sources like the Sun. It also
allows the simultaneous recording of dynamic radio spectra and images at
different frequencies. These capabilities make LOFAR a very powerful
instrument for imaging spectroscopy of celestial radio sources and to a radio
heliograph if applied to the Sun.

\section{The Solar Imaging Pipeline}
The Solar Imaging Pipeline is a software package for analyzing LOFAR
observations of the Sun. It was developed by the Solar KSP at the
Leibniz-Institut f\"ur Astrophysik Potsdam (AIP) as an extension of the LOFAR
Standard Imaging Pipeline \citep{2011JApA...32..589H} to accomplish a proper
data processing for solar LOFAR data. It is necessary because of the
differences between solar and standard observations and the consequences for
the data processing. They are summarized in
Table~\ref{table:SolarObservations} and \ref{table:DataProcessing} and
explained below.

\begin{table*}
  \caption[Differences between observations of the Sun and standard radio
    sources]{Differences between observations of the Sun and standard radio
    sources.} \centering
  \medskip
  \begin{tabular}{|c|C{4cm}|C{4cm}|}
    \hline
                                   & Standard radio source & Sun \\
    \hline
    Temporal resolution [s]        & $\ge 10^4$          & $\le 1$ \\
    Spatial resolution [arcsec]    & $\sim 1$            & $\sim 10^1$ \\
    No. of stations                & $\ge 48$            & $\le 30$ \\
    No. of baselines               & $> 10^3$            & $< 5 \times 10^2$\\
    Field of view [deg]            & $> 10^1$            & $\sim 1$ \\
    Typical flux density [Jy]      & $\ll 10^3$          & $10^4 - 10^{12}$ \\
    Typical noise level [Jy]       & $10^{-1}$           & $10^2$ \\
    \parbox[m]{5.5cm}{Dynamic range of flux density \\
      \centering [orders of magnitude]} & 0              & 8 \\
    Ionospheric scintillation      & low                 & high \\
    High res. dynamic spectroscopy & no                  & yes \\
    \hline
  \end{tabular}
  \label{table:SolarObservations}
  \bigskip

  \caption[Differences between standard and solar data processing]{Differences
    between standard and solar data processing resulting from the differences
    in the observations of Table~\ref{table:SolarObservations}.} \centering
  \medskip
  \begin{tabular}{|c|c|c|}
    \hline
                          & Standard Imaging        & Solar Imaging \\
    \hline
    Aperture synthesis    & yes                     & no \\
    Flagging              & in frequency \& time    & in frequency only \\
    uv-range [wavelengths]& $0-10^6$                & $0-10^3$ \\
    Sky model source threshold   & $\le 50$ Jy      & $\sim 10^3$ Jy \\
    Distance calibrator - target & $\le 1$ deg      & $\ge 5$ deg \\
    Demixing              & yes                     & no \\
    Imager                & AWImager                & CASA imager \\
    CLEAN algorithms & Clark for point sources &multi-scale for ext. sources\\
    Self-calibration      & yes                & limited to selected events\\
    Solar Data Center     & no                      & yes \\
    \hline
  \end{tabular}
  \label{table:DataProcessing}
\end{table*}

\subsection{Characteristics of solar observations}

\subsubsection{High variability of solar radio emission}
\begin{figure} \centering
    \includegraphics[width=\linewidth]{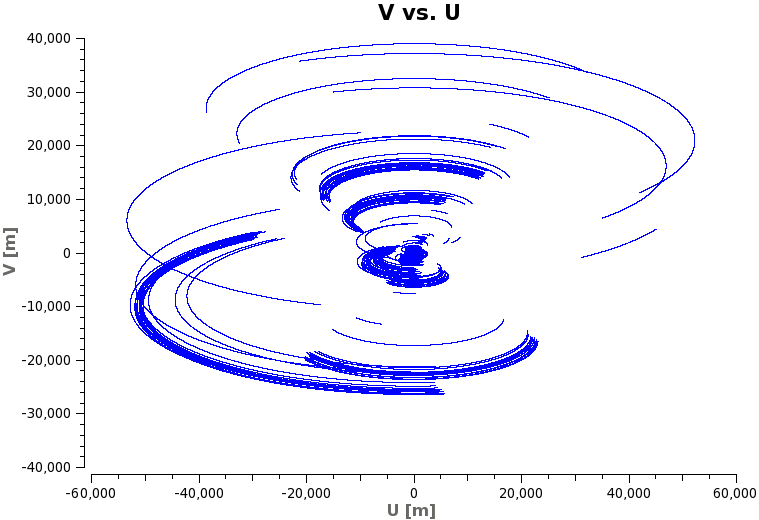}
    \caption[uv-coverage for 1 hour of data]{uv-coverage (baselines) for 12
      hours of data. Because of the Earth's rotation each point in the
      uv-plane multiplies into an arc like point set.}
    \label{fig:ApertureSynthesis}
\end{figure}

\begin{figure} \centering
    \includegraphics[width=\linewidth]{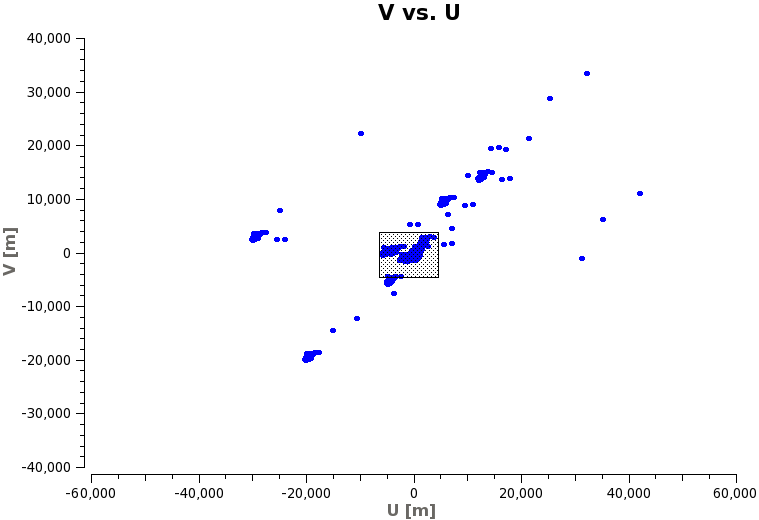}
    \caption[uv-coverage for 1 second of data]{uv-coverage for 1 second of
      data. The box indicates the approximate region of points (baselines)
      that remain if the uv-range is limited to $\le 1000$ wavelengths.}
    \label{fig:NoApertureSynthesis}
\end{figure}

The radio emission from most radio sources that LOFAR observes such as
galaxies is constant and ranges from 0.1 to 100~Jy. However, the emission from
the Sun at 100~MHz is about $10^4$~Jy (1 solar flux unit, sfu) during quiet
periods but it can rise to $10^{12}$~Jy ($10^8$ sfu) within a few seconds in
case of solar radio bursts \citep{2010LanB...4A..216M,2000GMS...119..115D}
which can occur anywhere near the Sun. So solar radio observations need to be
able to record a dynamic range in brightness of 8 orders of magnitude with a
time resolution of seconds. Consequences:

\begin{itemize}
\item A time resolution of seconds prevents the typical aperture synthesis
  which uses the Earth's rotation and observation times of hours for
  increasing the uv-coverage and image quality.
  Fig.~\ref{fig:ApertureSynthesis} and \ref{fig:NoApertureSynthesis} show the
  uv-coverage with and without aperture synthesis for comparison. Each point
  in the uv-plane represent a baseline.

\item The short duration of some bursts prevents temporal radio frequency
  interference (RFI) removal also called flagging, since flagging of radio
  peaks could also remove the signals of solar bursts.

\item Dynamic radio spectra are recorded simultaneously with the imaging data,
  since they can provide a higher time and frequency resolution at a lower
  data rate. They also support the identification of solar bursts in LOFAR
  data.
\end{itemize}

\subsubsection{Limited resolution in the corona}
While standard targets can be resolved with sub-arcsec resolution, the spatial
resolution of solar observations is limited to a few ten arcsec by turbulence
in the corona \citep{2004P&SS...52.1381B}. Consequences:

\begin{itemize}
\item Since the angular resolution is proportional to the inverse of the
  baseline length, the uv-range can be limited to a maximum baseline length of
  1000 wavelengths. The box in Fig.~\ref{fig:NoApertureSynthesis} marks the
  approximate subset of stations that remains with this limitation.

\item International stations and some remote stations are not needed since
  they provide only baselines with a length of more than 1000 wavelengths.
  This reduces the number of baselines $n_b = n_s \times (n_s-1) / 2$ (where
  $n_s$ is the number of stations) and therefore limits the image resolution.
\end{itemize}

\subsubsection{Sufficient field of view}
Since solar radio emission is a result of plasma emission determined by the
plasma density of the corona, radio emission in the frequency range from 10 to
240~MHz is limited to a region within 3$R_\odot$. This region is well
contained with the LOFAR field of view of approximately 5 degrees.
Consequence:

\begin{itemize}
\item A field of view of $3R_\odot$ does not require the corrections provided
  by the AWImaging \citep{2013A&A...553A.105T} and so the CASA imager can be
  used as well. This was an advantage early in LOFAR's commissioning phase
  when the AWImager was not available and when functionality is required that
  is not yet implemented.
\end{itemize}

\begin{figure*}\centering
  \includegraphics[width=\linewidth]{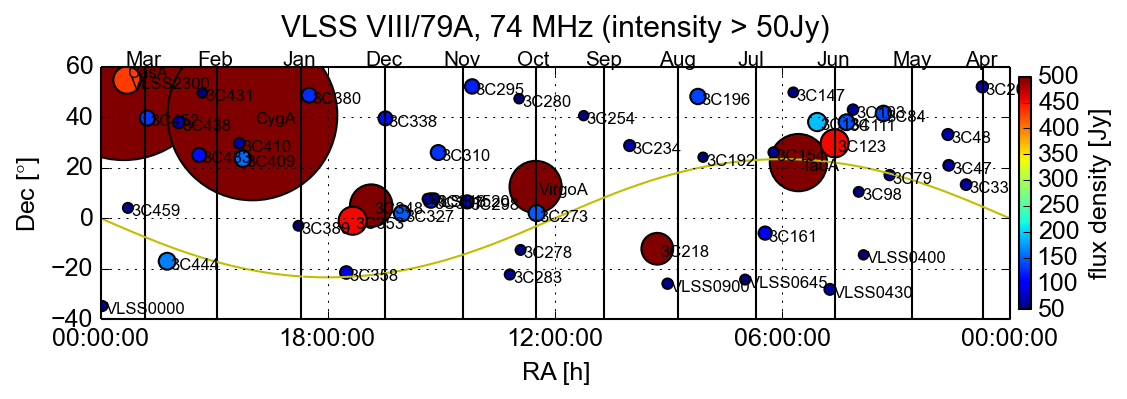}
    \caption[Sky map of intense radio sources]{Sky map of intense radio
      sources and their location with respect to the Sun. The yellow line
      shows the path of the Sun during one year. The area of each source
      represents its flux density. A compact source with a high flux density
      and a distance of a few 10 degrees is best suited for the calibration of
      solar data.}
  \label{fig:Calibrator}
\end{figure*}

\begin{figure*}\centering
  \includegraphics[width=\linewidth]{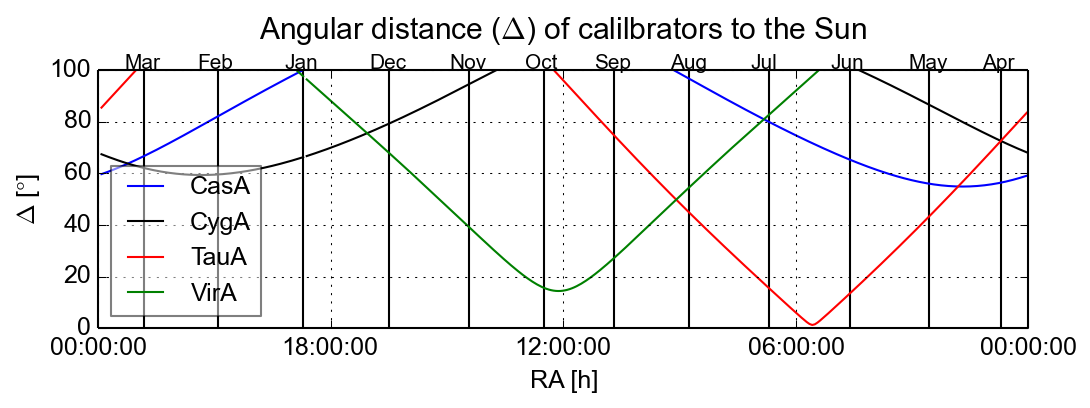}
  \caption[Distance of solar calibrators]{Annual variation of the distance
    ($\Delta$) of sources with high flux density that are used as calibrator
    for solar LOFAR data.}
  \label{fig:Calibrator-dist}
\end{figure*}

\begin{figure}\centering
  \includegraphics[width=\linewidth]{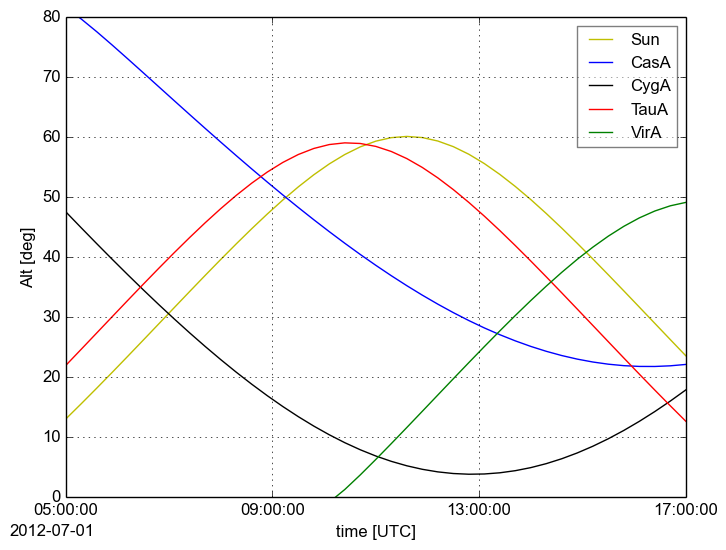}
  \caption[Altitude of solar calibrators]{Daily variation of the altitude of
    solar calibrators. Low altitude observations provide data of lower quality
    which are also harder to calibrate. The LSDC provides these plots for
    every day of the year to support the scheduling of observations.}
  \label{fig:Calibrator-alt}
\end{figure}

\begin{figure*} \centering
  \includegraphics[width=\linewidth]{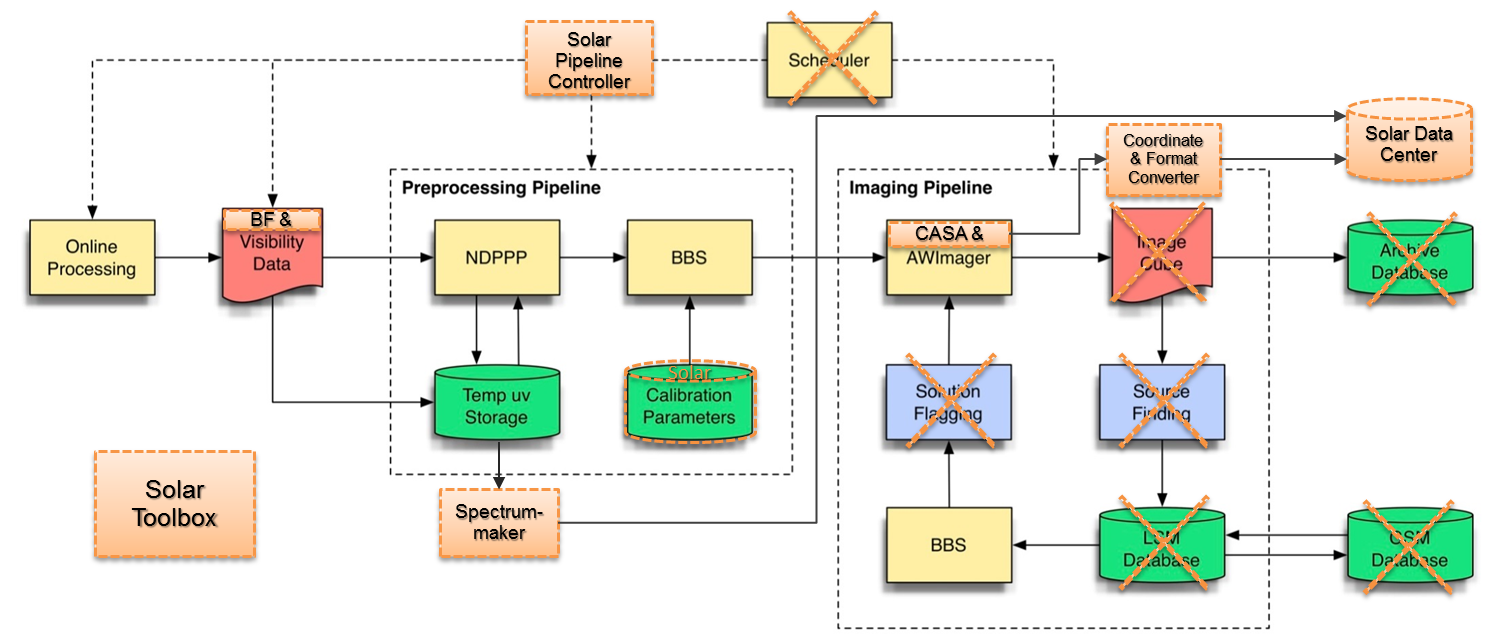}
  \caption[Flowchart of the Standard and Solar Imaging Pipeline]{Flowchart of
    the Standard Imaging Pipeline and the modifications (orange) that lead to
    the Solar Imaging Pipeline. The data flow starts with the online
    processing and ends at the Solar Data Center. The crossed out components
    are not needed for solar imaging. Details of the individual components are
    described in the text.}
  \label{fig:SolarImagingPipeline}
\end{figure*}

\subsubsection{High flux density of solar radio emission}
Another difference exists with the radio sources that can be used for
calibration. For many sky regions detailed sky models exist with a resolution
of a few Jy which the Sun passes during a year. But since the Sun emits a flux
density of $10^4$~Jy or orders of magnitudes more its radiation dominates the
signal from these sky regions and is picked up through LOFAR's side lobes.
Consequences:

\begin{itemize}
\item Only strong radio sources can serve as calibrators. These sources are
  Cassiopeia A, Cygnus A, Taurus A and Virgo A (Fig.~\ref{fig:Calibrator}).
  For this reason solar observations are done with two sub-array pointings.
  One is directed at the Sun and the other at a suitable calibrator source.
  This way the calibration coefficients can be obtained from the calibrator
  data and then applied to the data from the Sun. However, this adds
  additional constraints for the observation. First the angular distance
  between the Sun and the calibrator may neither be too small ($< 5$~deg) nor
  too large ($> 50$~deg), since otherwise the calibrator will be perturbed by
  the Sun or the calibration will be perturbed by an imperfect beam model,
  respectively. Fig.~\ref{fig:Calibrator-dist} shows the distance ($\Delta$)
  of the calibrators to the Sun versus time. Second the Sun and the calibrator
  need to be observed at sufficiently high altitude angles ($> 20$~deg) since
  otherwise their radio emission is too much perturbed by the atmosphere. This
  condition also excludes observations during winter months from November to
  February. Fig.~\ref{fig:Calibrator-alt} shows the altitudes which have to be
  taken into account for an observation on July~1.

\item Similar to the emission from the Sun also the emission from the
  calibrators effects the solar data. Therefore its influence should be
  removed by demixing. Demixing is a method in which the contribution from
  certain radio sources is modeled and subtracted from the data.
  Unfortunately, demixing often does not work well without a model for the
  radio emission from the Sun which is not available at pre-processing, when
  demixing is applied for standard imaging. So demixing is not yet part of
  imaging of the Sun.
\end{itemize}

\subsubsection{Observations during day time}
LOFAR Solar observations have to be done during daytime, preferably at noon.
At noon ionospheric scintillation also reach their maximum. Consequence:

\begin{itemize}
\item Ionospheric scintillation causes different phase delays for different
  atmospheric regions the radio signals pass on their way to different LOFAR
  stations. This leads to calibration and imaging errors, more severely during
  day time than during night time. This is also another reason for a lower
  image quality of solar observations than of other observations which are
  usually taken during night time.
\end{itemize}

\subsection{The solar data processing}

The solar imaging pipeline was developed according to the above criteria. The
outline is shown in Fig.~\ref{fig:SolarImagingPipeline}. It shows the data
flow and in orange color the modifications from the Standard Imaging Pipeline.
The individual steps and components are discussed below.

\subsubsection{Online processing}
The LOFAR data from all stations is correlated at a computer cluster in
Groningen. This correlates the data in real-time and stores it at a temporary
storage for a later processing. Currently either low or high-band antenna data
is recorded at several equidistant frequencies. Each of these frequency
subbands has a bandwidth of 195 kHz and a subdivision into 64 channels. During
solar observations also beam-formed data for dynamic spectroscopy is recorded
in addition to visibility data which is used for standard imaging.

\subsubsection{Temporary storage and data formats}
The immediate processing of the data as shown in
Fig.~\ref{fig:SolarImagingPipeline} is not yet realized but a planned
processing option for the future. Instead the data is currently sent to a
temporary storage. The imaging and spectroscopic data is recorded in CASA
Measurement Sets (MS, \cite{2007ASPC..376..127M}) and LOFAR's beam-formed
format \citep{2011A&A...530A..80S} with an HDF5 container \citep{HDF5}
respectively. An MS contains the data of only one subband. The raw data rate
scales with the number of channels recorded and is roughly 30 times larger
than data with a single channel average. The typical size for the latter is
about 1~GB for 10 hours of data with time resolution of 1s. The corresponding
JPEG images and thumbnails have about the same size, while the FITS images
need about three times as much. The size of beam-formed data scales with the
time resolution and number of channels. It is similar to the size of one MS
while providing a higher resolution. In summary a typical solar observation of
a few hours produces a few TB of data.

\subsubsection{New Default Pre-Processing Pipeline (NDPPP)}
The first step in the pre-processing of imaging data is done with the New
Default Pre-Processing Pipeline (NDPPP). This includes flagging of the
channels of the calibrator data for RFI with the AOFlagger
\citep{2013A&A...549A..11O}. The Sun data are not flagged, to preserve all
variable radio features. Subsequently, the channels (excluding edge channel)
of all subbands are averaged to one channel per subband and the uv-range is
limited to 0-1000 wavelengths.

\subsubsection{BlackBoard Selfcal (BBS)}
The second step of the pre-processing is the calibration with LOFAR's
BlackBoard Selfcal (BBS) system. Here amplitude and phase corrections are
determined for every 30 seconds of calibrator data using a corresponding sky
model. The solutions are then transferred to the data from the Sun.
Alternatively also two Gaussian sky models of the Sun for low- and high-bands
have been developed which can be used for direct calibration of the data from
the Sun. However this calibration is less accurate because of the extension of
the model only short baselines can be calibrated. Also the determination of
the astrometric position is limited to a few arcmin because of the asymmetric
variability of the corona.

\subsubsection{Solar calibration parameters}
BBS is controlled through certain configuration parameters. The specific
parameters for the Sun are derived from the standard imaging parameters and
are included in the Solar Imaging Pipeline. They are optimized for the
calibration of strong calibrator sources and a transfer of the calibration
coefficients to the solar data.

\subsubsection{Imaging}
After the calibration the corrected data is imaged with the multi-scale CLEAN
algorithm provided by CASA \citep{2007ASPC..376..127M}. The final image is
then transformed from equatorial to solar coordinates where the Sun is
centered and the Sun's north pole is oriented upwards. This is done for every
time step and frequency. Alternatively the CASA imager can be replaced by the
AWImager. The AWImager provides additional correction by the w- and
A-projection. The former projection removes the effects of non co-planar
baselines in large fields. The latter takes into account a varying primary
beam during synthesis observations. However, both corrections are not
necessary for the small field of view used in solar observations.

With the CASA imager also images from multiple integration steps can be
combined. This so-called aperture synthesis virtually increases the number of
baselines by combining data from different times where the antenna positions
have shifted because of the Earth's rotation. Aperture synthesis for the Sun
is complicated since its RA/Dec coordinates change while imaging is normally
done for a fixed RA/Dec coordinate. Therefore the aperture synthesis was
extended for sources with changing RA/Dec coordinates. It can be activated via
the ``movingsource'' option of the CASA imager. With the key word ``SUN''
every time step of an observation is shifted such that the position of the Sun
remains fixed. This avoids smearing of the Sun during longer observations
periods of hours.

\subsubsection{Self-calibration}
Optionally an additional phase-only self-calibration can be applied to
selected images. In the standard imaging pipeline a source finding algorithm
is applied to the image to find sources not contained in the sky model. These
are then added to the sky model and applied in a new calibration. For solar
observations a simplified method is used which directly uses the cleaned
images as model. Since it does not require assumptions about the sources to
find, it is useful where little is known about the radio source morphology.
Then imaging is repeated resulting in a reduced noise and increased source
brightness. Further self-calibration cycles can be repeated if needed until no
further improvement is achieved. This is usually the case after one or two
cycles.

\subsubsection{Simulation of images}
\label{sec:simulations}
\begin{figure}
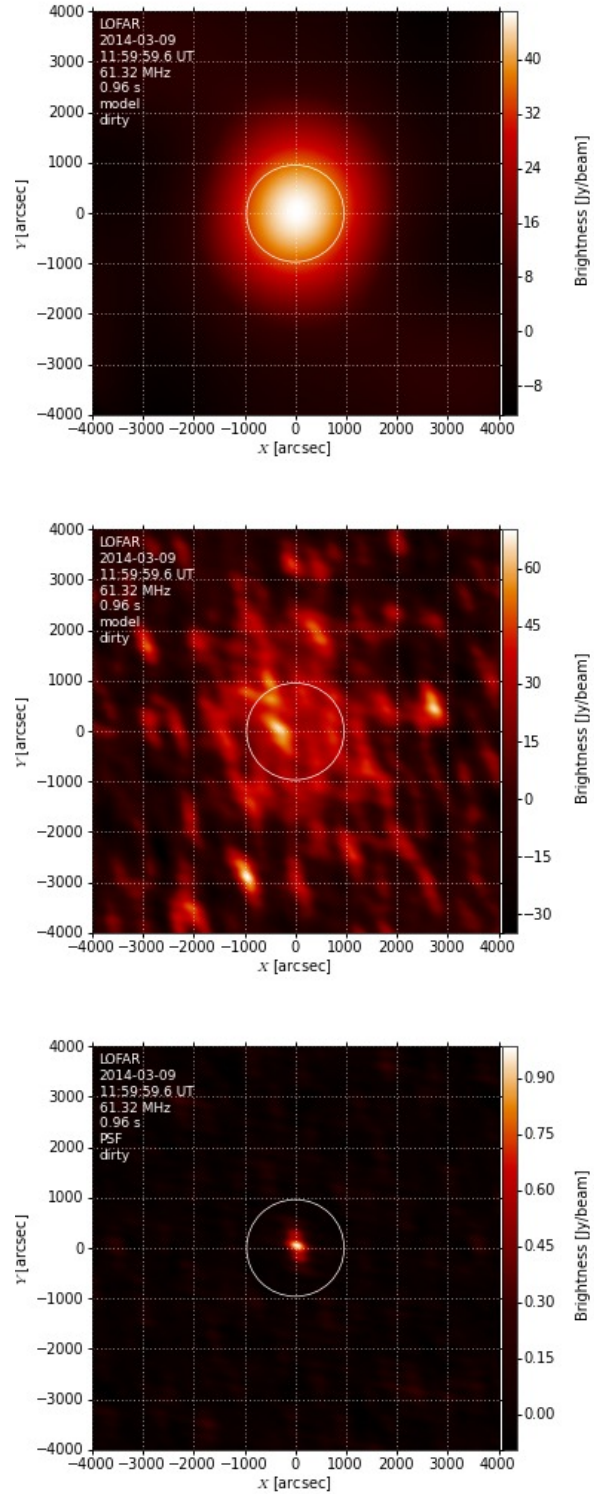
\centering
  \includegraphics[width=\linewidth]{{{1_20140309T115959.66-45025_model}}}
  \includegraphics[width=\linewidth]{{{1_20140309T115959.65-00001_model+Sun+Ateam}}}
  \includegraphics[width=\linewidth]{{{1_20140309T115959.66-45025_psf}}}
    \caption[Simulated radio images]{Simulated radio images of a
      two-dimensional Gaussian model of the Sun (top), this model with
      side-lobe contributions from the four strong radio sources mentioned in
      section~\ref{sec:simulations} (middle) and of the PSF (bottom) for an
      observation on March 9, 2014.}
  \label{fig:Simulations}
\end{figure}

The solar imaging pipeline automatically simulates images of a Gaussian model
of the Sun, the point spread function (PSF, Fig.~\ref{fig:Simulations}) and
the contributions from the intense radio sources Cygnus A, Cassiopeia A,
Taurus A and Virgo A through side-lobes to the model. This is very instructive
for understanding the solar images and their fidelity for a given array
configuration. The simulated images are accessible through the LOFAR Solar
Data Center described in section~\ref{sec:LSDC}.

\subsubsection{Solar Pipeline controller}
The Solar Imaging Pipeline controller replaces the scheduler of the standard
imaging pipeline. It is capable of the data management, which includes the
distribution of data to the LOFAR processing cluster and its collection. It
also controls the steps of the solar imaging pipeline the LOFAR data is run
through.

\subsubsection{Solar toolbox}
The solar toolbox contains further programs for various tasks related to the
data processing. This includes calculations of solar coordinates and angles,
image format conversion, extraction of specific observation periods, job and
disk space monitoring, the correction of pointing positions and many other
tasks.

\begin{figure*} \centering
  \includegraphics[width=\linewidth]{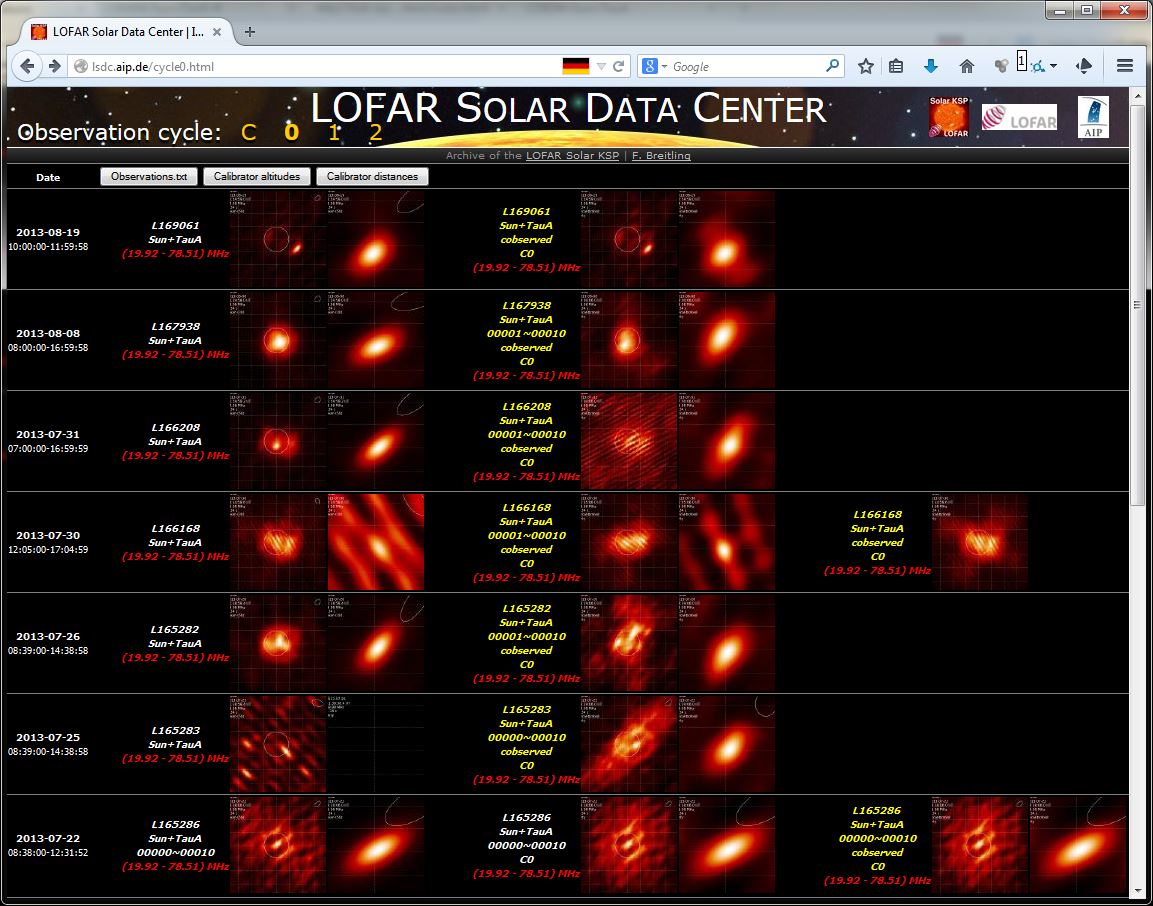}
  \caption[Main page of the LOFAR Solar Data Center]{Main page of the LOFAR
    Solar Data Center showing data from observations cycle 0. The available
    observations are listed chronologically line by line and linked to their
    corresponding observation page.}
  \label{fig:LSDC}
\end{figure*}

\subsubsection{Dynamic spectra}
The Solar Imaging Pipeline also provides the spectrum maker for processing
beam-formed data. This identifies and excludes channels with a high tenth
percentile caused by RFI. It also excludes channels close to the edge of a
subband which contain data of low quality. It averages the remaining channels
of each subband and scales them to values between 0 and 1. To make the scaling
more robust against RFI the highest 1\% of the recorded values are ignored for
the scaling and truncated to 1. To emphasize the burst signatures the values
are gamma corrected. Intermediate values between the different subbands are
interpolated.

\section{The LOFAR Solar Data Center}
\label{sec:LSDC}
The LOFAR Solar Data Center (LSDC) is the archive for LOFAR data products of
the Solar KSP. Its software is part of the Solar Imaging Pipeline and it is
located and operated at the AIP. It provides a web user interface at
\url{http://lsdc.aip.de/} to share LOFAR solar observations with the solar
science community. It was designed for browsing through available data, for
finding specific observations and for presenting the data products in a way
that supports the identification of solar events and their interpretation. To
serve this purpose the LSDC is structured as follows.

\begin{figure*} \centering
  \includegraphics[width=\linewidth]{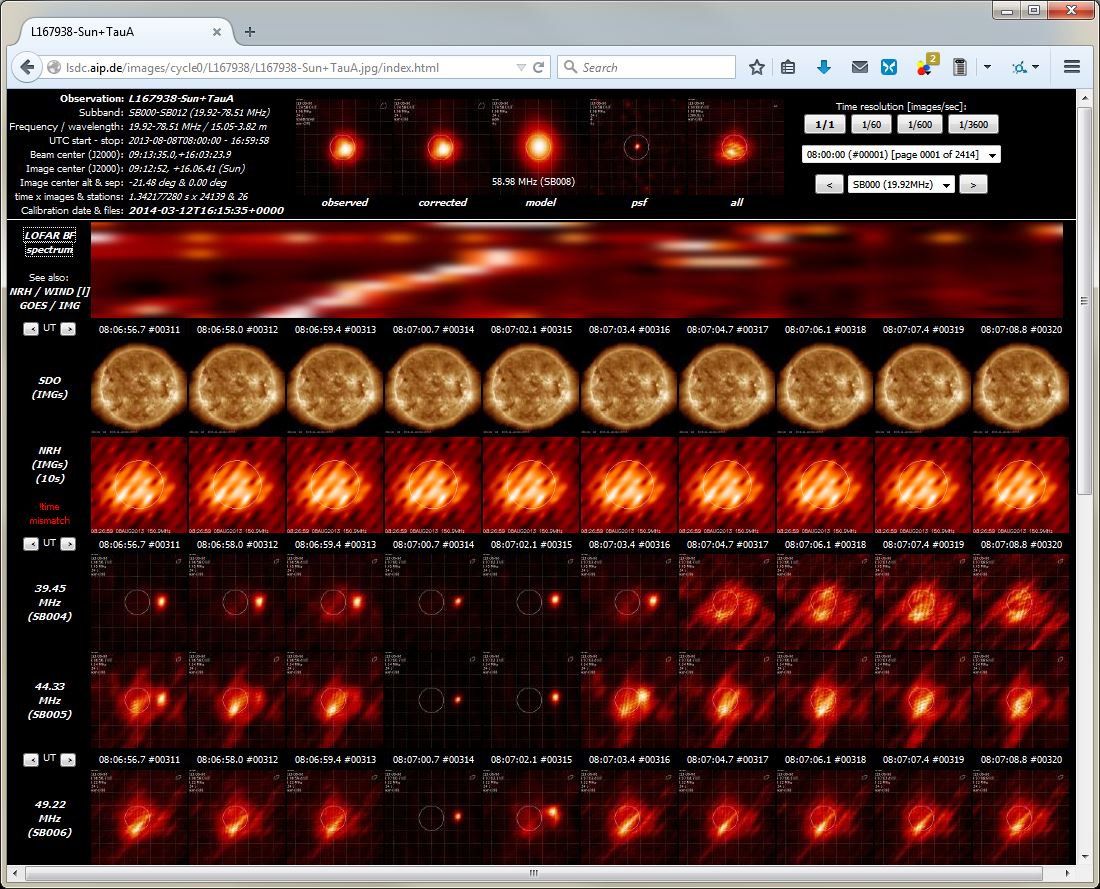}
  \caption[Observation Page]{LSDC page of observation L206894 providing
    observation details, diagnostic images and a navigation box in the first
    row. Spectra, images from other instruments and images of the individual
    subbands follow. All images are previews that are linked to their full
    resolution version. A few solar type III bursts are visible in the dynamic
    radio spectrum and in the corresponding images, which also show their
    location.}
  \label{fig:LSDC-obspage}
\end{figure*}

\begin{figure*} \centering
  \includegraphics[width=\linewidth]{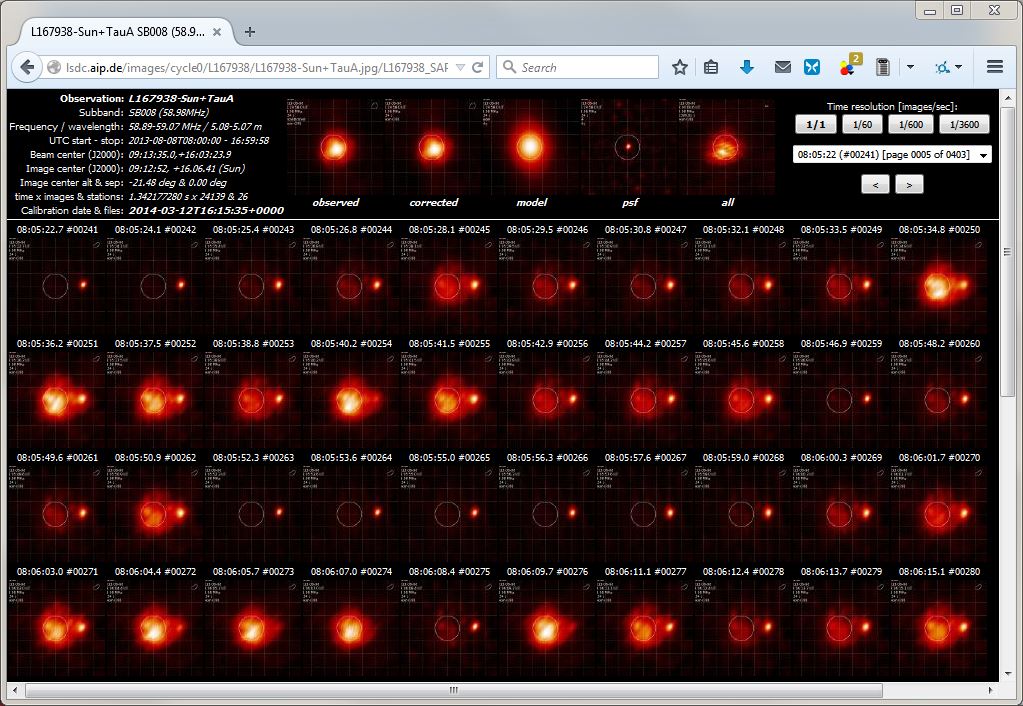}
  \caption[Subband page]{The page of subband 180 (65~MHz) of observation
    L206894. It can show longer periods of a single subband than the
    observation page. Here the sudden brightening within seconds in the north
    east reveals a type III burst. The first row is very similar to the first
    row of the observation page and has the same functionality.}
  \label{fig:LSDC-sbpage}
\end{figure*}

\begin{figure*} \centering
  \includegraphics[width=\linewidth]{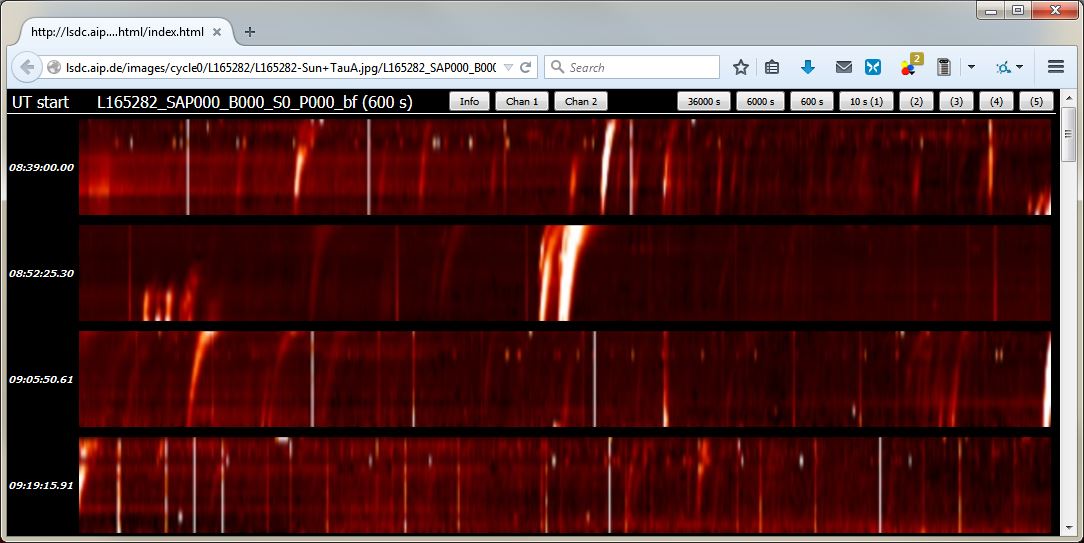}
  \caption[Dynamic spectrum page]{The LOFAR dynamic spectrum page is useful to
    identify solar events such as solar radio bursts, which are clearly
    visible as drifting broadband peaks. The top row provides further
    information and functionality explained in the text.}
  \label{fig:LSDC-spectra}
\end{figure*}

\subsection{Main page}
The LSDC is entered through the main page shown in Fig.~\ref{fig:LSDC}. It
shows a header with the title, icons and a menu at the left for switching
between different observation cycles to be displayed in the main section of
the page. Currently available options are \texttt{C} for the commissioning
cycle and \texttt{0} to \texttt{2} for the regular operation cycles. Here
cycle 0 is selected for the further discussion. The main section starts with a
row of three buttons. They link to the list of available observations and to
plots of the altitudes of the solar calibration sources
(Fig.~\ref{fig:Calibrator-alt}) and their distances to the Sun
(Fig.~\ref{fig:Calibrator-dist}). The plots provide useful information for
selecting calibrators for observations and the data processing. Next a
chronological table follows with the available data of the observation cycle.
Each row represents one observation and its available data products. It starts
with the observation date including start and end time followed by the results
from different processing strategies. Each result contains information about
the observation ID, the target and calibrator source, possible additional
information regarding the calibration and cleaning parameters, the observation
frequency and a preview images of the Sun and the calibrator. Each image links
to the corresponding observation page with all related data products.

\subsection{Observation page}
The observation pages provide access to the available data products for
viewing and downloading. It also provides a control box to browse through
them. An example is Fig.~\ref{fig:LSDC-obspage} which represents observation
L206894. It is organized in rows, each showing the image sequence of a
subband. However the first four rows are different:

\begin{enumerate}
\item The first row starts with a summary of observation specifications and a
  link to configuration and logfiles from the processing. The following five
  diagnostic images give an idea about the data and calibration quality. They
  show the un-calibrated data, the calibrated data, a simulated model and the
  point spread function (PSF) for the time step at the middle of the
  observation. The fifth image is integrated over the complete observation
  period. The row ends with the control box for browsing through the data. It
  starts with four buttons to select different time resolutions that shows
  every first, 60th, 600th and 3600th image. The drop-down menu in the next
  line lets the user jump directly to a specific time range. The arrow buttons
  in the line below steps back or forward in time. The last drop-down menu in
  this last line opens the page of a selected subband in a new window.

\item The second row shows the dynamic radio spectrum by LOFAR. It also
  provides links to dynamic spectra from the Nancay Decameter Array
  \citep{2000GMS...119..321L}, the WIND satellite \citep{1995SSRv...71..231B}
  and the Geostationary Operational Environmental Satellites (GOES).

\item The third row serves as a time axis showing the image numbers and
  recording times. It also contains navigation buttons to step one page back
  or forward. It repeats after every two image rows.

\item The fourth row shows extreme ultraviolet images of the lower corona from
  SDO \citep{2012SoPh..275...17L} with a time resolution of minutes.

\item The fifth row shows the images from the Nancay Radio Heliograph
  \citep{1997LNP...483..192K} at 151~MHz with a time resolution of 10~s.

\item The sixth row shows the same time axis as the third row.

\item From the seventh row on the LOFAR images from the different subbands
  follow with increasing frequency.
\end{enumerate}

The images are only previews linked to the full resolution images. The page
also contains links to the individual subband pages and to the dynamic
spectrum page.

\subsection{Subband page}
The subband page shows images of a single subband and so can display a longer
image sequence. This is useful for studying the temporal evolution of a solar
radio event over longer period. Fig.~\ref{fig:LSDC-sbpage} shows subband 180
(65~MHz) of observation L206894 as an example. The first row is very similar
to the first row of the observation page and provides the same functionality.

\subsection{Dynamic spectrum page}

\begin{figure} \centering
  \includegraphics[width=\linewidth]{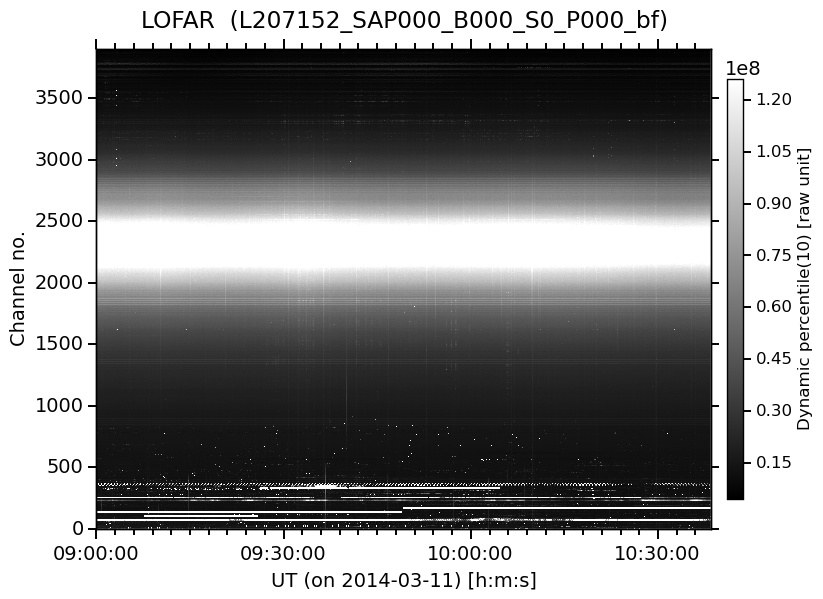}
  \caption[Dynamic spectrum of the channels of the LOFAR beam-formed
    data]{Dynamic spectrum of the channels of LOFAR beam-formed data. This is
    useful to check the data for RFI.}
  \label{fig:DynamicChannelRFI}
\end{figure}

\begin{figure} \centering
  \includegraphics[width=.9\linewidth]{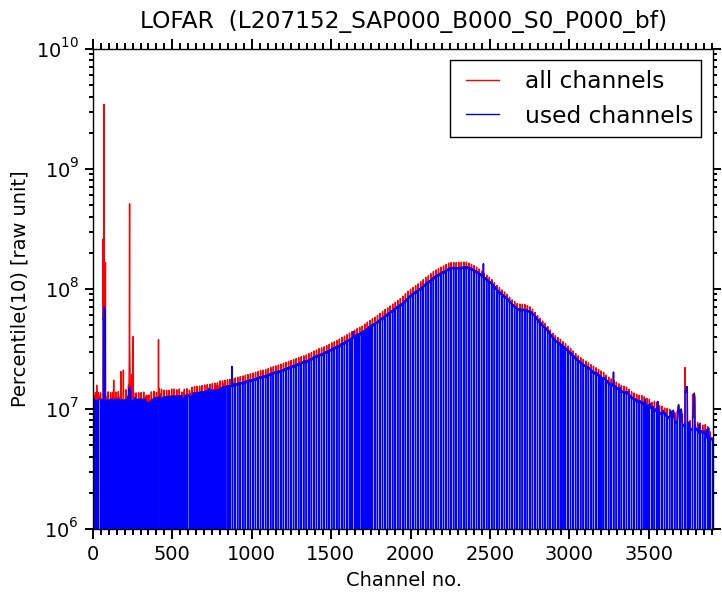}
  \caption[Channel values of LOFAR beam-formed data]{Channel values of LOFAR
    beam-formed data. Channels with values within the tenth percentile of one
    subband and channels close to the edge of a subband (red) are flagged and
    removed. The remaining channels (blue) are used to build the LOFAR dynamic
    spectrum.}
  \label{fig:ChannelRFI}
\end{figure}

The dynamic spectrum page is shown in Fig.~\ref{fig:LSDC-spectra}. Its purpose
is to display the LOFAR dynamic spectra next to each other to cover a long
time range. This arrangement is very useful for identifying solar radio
bursts. Each row continues the time range covered by the previous row. Each
spectrum is only a preview and links to its full resolution version. The first
row shows the observation id and buttons with additional functionality. One
provides further information about the observation settings, two others show
RFI (Fig.~\ref{fig:DynamicChannelRFI} and \ref{fig:ChannelRFI}), some can
change the time resolution of the spectra and others can select pages of a
certain time range.

\section{Conclusion}
LOFAR is a new radio interferometer for dynamic imaging spectroscopy that can
be used for observations of the Sun. Solar observations are managed by the
LOFAR Key Science Project ``Solar Physics and Space Weather with LOFAR'',
which has developed the Solar Imaging Pipeline and the Solar Data Center for
processing the LOFAR data and providing access to the data product by the
solar physics community. The Solar Imaging Pipeline builds upon the Standard
Imaging Pipeline and differs in several ways as required by solar radio
observations. The Solar Data Center is the archive of the Solar KSP and is
located at the AIP. It provides access to processed data from solar
observations through a web interface. The LOFAR spectra and images of the Sun
have been compared with other observations by other instruments such as the
Nancay Radio Heliograph and the Nancay Decameter Array
\citep{2000GMS...119..321L} and good agreement has been found. The successful
verification permits the usage of solar LOFAR observations for scientific
studies, which will be discussed in forthcoming publications.

\section*{Acknowledgement}
The authors would like to thank A. Warmuth, H. \"Onel and the LOFAR builders
in particular the LOFAR software developers G. v. Diepen, S. Duscha, R.
Fallows, G. Heald, R. Pizzo, C. Tasse, N. Vilchey, R. v. Weeren, M. Wise, J.
v. Zwieten for supporting the development of the Solar Imaging Pipeline with
useful discussions and advice; the CASA support and developers for essential
implementations to CASA with respect to solar imaging; their colleagues H.
Enke, J. Klar and A. Khalatyan from the AIP e-Science section for their
support in providing the data storage, computer cluster and web server for
the LOFAR Solar Data Center. Financial support was provided by the German
Federal Ministry of Education and Research (BMBF in the framework of the
Verbundforschung, D-LOFAR 05A11BAA). LOFAR, designed and constructed by
ASTRON, has facilities in several countries that are owned by various parties
(each with their own funding sources) and are collectively operated by the
International LOFAR Telescope (ILT) foundation under a joint scientific
policy.

\ifbibtex
\section*{References}
\bibliographystyle{hapj}
\bibliography{bibtex/solar,bibtex/lofar}

\begin{thebibliography}{23}
\expandafter\ifx\csname natexlab\endcsname\relax\def\natexlab#1{#1}\fi

\bibitem[{{Bastian}(2004)}]{2004P&SS...52.1381B}
{Bastian}, T.~S. 2004, \planss, 52, 1381

\bibitem[{{Bougeret} {et~al.}(1995){Bougeret}, {Kaiser}, {Kellogg}, {Manning},
  {Goetz}, {Monson}, {Monge}, {Friel}, {Meetre}, {Perche}, {Sitruk}, \&
  {Hoang}}]{1995SSRv...71..231B}
{Bougeret}, J.-L. {et~al.} 1995, \ssr, 71, 231

\bibitem[{{Breitling} {et~al.}(2011){Breitling}, {Mann}, \&
  {Vocks}}]{2011pre7.conf..373B}
{Breitling}, F., {Mann}, G., \& {Vocks}, C. 2011, Planetary, Solar and
  Heliospheric Radio Emissions (PRE VII), 373, 1511.03123

\bibitem[{{Dulk}(2000)}]{2000GMS...119..115D}
{Dulk}, G.~A. 2000, Washington DC American Geophysical Union Geophysical
  Monograph Series, 119, 115

\bibitem[{{Heald} {et~al.}(2011){Heald}, {Bell}, {Horneffer}, {Offringa},
  {Pizzo}, {van der Tol}, {van Weeren}, {van Zwieten}, {Anderson}, {Beck}, {van
  Bemmel}, {B{\^i}rzan}, {Bonafede}, {Conway}, {Ferrari}, {de Gasperin},
  {Haverkorn}, {Jackson}, {Macario}, {McKean}, {Miraghaei}, {Orr{\`u}},
  {Rafferty}, {R{\"o}ttgering}, {Scaife}, {Shulevski}, {Sotomayor}, {Tasse},
  {Trasatti}, \& {Wucknitz}}]{2011JApA...32..589H}
{Heald}, G. {et~al.} 2011, Journal of Astrophysics and Astronomy, 32, 589,
  1106.3195

\bibitem[{{Kerdraon} \& {Delouis}(1997)}]{1997LNP...483..192K}
{Kerdraon}, A., \& {Delouis}, J.-M. 1997, in Lecture Notes in Physics, Berlin
  Springer Verlag, Vol. 483, Coronal Physics from Radio and Space Observations,
  ed. G.~{Trottet}, 192

\bibitem[{{Lecacheux}(2000)}]{2000GMS...119..321L}
{Lecacheux}, A. 2000, Washington DC American Geophysical Union Geophysical
  Monograph Series, 119, 321

\bibitem[{{Lemen} {et~al.}(2012){Lemen}, {Title}, {Akin}, {Boerner}, {Chou},
  {Drake}, {Duncan}, {Edwards}, {Friedlaender}, {Heyman}, {Hurlburt}, {Katz},
  {Kushner}, {Levay}, {Lindgren}, {Mathur}, {McFeaters}, {Mitchell}, {Rehse},
  {Schrijver}, {Springer}, {Stern}, {Tarbell}, {Wuelser}, {Wolfson}, {Yanari},
  {Bookbinder}, {Cheimets}, {Caldwell}, {Deluca}, {Gates}, {Golub}, {Park},
  {Podgorski}, {Bush}, {Scherrer}, {Gummin}, {Smith}, {Auker}, {Jerram},
  {Pool}, {Soufli}, {Windt}, {Beardsley}, {Clapp}, {Lang}, \&
  {Waltham}}]{2012SoPh..275...17L}
{Lemen}, J.~R. {et~al.} 2012, \solphys, 275, 17

\bibitem[{{Mann}(2006)}]{2006GMS...165..221M}
{Mann}, G. 2006, Washington DC American Geophysical Union Geophysical Monograph
  Series, 165, 221

\bibitem[{{Mann}(2010)}]{2010LanB...4A..216M}
------. 2010, Landolt B{\"o}rnstein, 216

\bibitem[{{Mann} {et~al.}(1999){Mann}, {Jansen}, {MacDowall}, {Kaiser}, \&
  {Stone}}]{1999A&A...348..614M}
{Mann}, G., {Jansen}, F., {MacDowall}, R.~J., {Kaiser}, M.~L., \& {Stone},
  R.~G. 1999, \aap, 348, 614

\bibitem[{{Mann} {et~al.}(2011){Mann}, {Vocks}, \&
  {Breitling}}]{2011pre7.conf..507M}
{Mann}, G., {Vocks}, C., \& {Breitling}, F. 2011, Planetary, Solar and
  Heliospheric Radio Emissions (PRE VII), 507

\bibitem[{{McLean}(1985)}]{1985srph.book...37M}
{McLean}, D.~J. 1985, in Solar Radiophysics: Studies of Emission from the Sun
  at Metre Wavelengths, ed. D.~J. {McLean} \& N.~R. {Labrum} (Cambridge and New
  York, Cambridge University Press), 37--52

\bibitem[{{McMullin} {et~al.}(2007){McMullin}, {Waters}, {Schiebel}, {Young},
  \& {Golap}}]{2007ASPC..376..127M}
{McMullin}, J.~P., {Waters}, B., {Schiebel}, D., {Young}, W., \& {Golap}, K.
  2007, in Astronomical Society of the Pacific Conference Series, Vol. 376,
  Astronomical Data Analysis Software and Systems XVI, ed. R.~A. {Shaw},
  F.~{Hill}, \& D.~J. {Bell}, 127

\bibitem[{{Melrose}(1985)}]{1985srph.book..177M}
{Melrose}, D.~B. 1985, in Solar Radiophysics: Studies of Emission from the Sun
  at Metre Wavelengths, ed. D.~J. {McLean} \& N.~R. {Labrum} (Cambridge and New
  York, Cambridge University Press), 177--210

\bibitem[{{Offringa} {et~al.}(2013){Offringa}, {de Bruyn}, {Zaroubi}, {van
  Diepen}, {Martinez-Ruby}, {Labropoulos}, {Brentjens}, {Ciardi}, {Daiboo},
  {Harker}, {Jeli{\'c}}, {Kazemi}, {Koopmans}, {Mellema}, {Pandey}, {Pizzo},
  {Schaye}, {Vedantham}, {Veligatla}, {Wijnholds}, {Yatawatta}, {Zarka},
  {Alexov}, {Anderson}, {Asgekar}, {Avruch}, {Beck}, {Bell}, {Bell}, {Bentum},
  {Bernardi}, {Best}, {Birzan}, {Bonafede}, {Breitling}, {Broderick},
  {Br{\"u}ggen}, {Butcher}, {Conway}, {de Vos}, {Dettmar}, {Eisloeffel},
  {Falcke}, {Fender}, {Frieswijk}, {Gerbers}, {Griessmeier}, {Gunst},
  {Hassall}, {Heald}, {Hessels}, {Hoeft}, {Horneffer}, {Karastergiou},
  {Kondratiev}, {Koopman}, {Kuniyoshi}, {Kuper}, {Maat}, {Mann}, {McKean},
  {Meulman}, {Mevius}, {Mol}, {Nijboer}, {Noordam}, {Norden}, {Paas}, {Pandey},
  {Pizzo}, {Polatidis}, {Rafferty}, {Rawlings}, {Reich}, {R{\"o}ttgering},
  {Schoenmakers}, {Sluman}, {Smirnov}, {Sobey}, {Stappers}, {Steinmetz},
  {Swinbank}, {Tagger}, {Tang}, {Tasse}, {van Ardenne}, {van Cappellen}, {van
  Duin}, {van Haarlem}, {van Leeuwen}, {van Weeren}, {Vermeulen}, {Vocks},
  {Wijers}, {Wise}, \& {Wucknitz}}]{2013A&A...549A..11O}
{Offringa}, A.~R. {et~al.} 2013, \aap, 549, A11, 1210.0393

\bibitem[{{Stappers} {et~al.}(2011){Stappers}, {Hessels}, {Alexov}, {Anderson},
  {Coenen}, {Hassall}, {Karastergiou}, {Kondratiev}, {Kramer}, {van Leeuwen},
  {Mol}, {Noutsos}, {Romein}, {Weltevrede}, {Fender}, {Wijers}, {B{\"a}hren},
  {Bell}, {Broderick}, {Daw}, {Dhillon}, {Eisl{\"o}ffel}, {Falcke},
  {Griessmeier}, {Law}, {Markoff}, {Miller-Jones}, {Scheers}, {Spreeuw},
  {Swinbank}, {Ter Veen}, {Wise}, {Wucknitz}, {Zarka}, {Anderson}, {Asgekar},
  {Avruch}, {Beck}, {Bennema}, {Bentum}, {Best}, {Bregman}, {Brentjens}, {van
  de Brink}, {Broekema}, {Brouw}, {Br{\"u}ggen}, {de Bruyn}, {Butcher},
  {Ciardi}, {Conway}, {Dettmar}, {van Duin}, {van Enst}, {Garrett}, {Gerbers},
  {Grit}, {Gunst}, {van Haarlem}, {Hamaker}, {Heald}, {Hoeft}, {Holties},
  {Horneffer}, {Koopmans}, {Kuper}, {Loose}, {Maat}, {McKay-Bukowski},
  {McKean}, {Miley}, {Morganti}, {Nijboer}, {Noordam}, {Norden}, {Olofsson},
  {Pandey-Pommier}, {Polatidis}, {Reich}, {R{\"o}ttgering}, {Schoenmakers},
  {Sluman}, {Smirnov}, {Steinmetz}, {Sterks}, {Tagger}, {Tang}, {Vermeulen},
  {Vermaas}, {Vogt}, {de Vos}, {Wijnholds}, {Yatawatta}, \&
  {Zensus}}]{2011A&A...530A..80S}
{Stappers}, B.~W. {et~al.} 2011, \aap, 530, A80, 1104.1577

\bibitem[{{Suzuki} \& {Dulk}(1985)}]{1985srph.book..289S}
{Suzuki}, S., \& {Dulk}, G.~A. 1985, in Solar Radiophysics: Studies of Emission
  from the Sun at Metre Wavelengths, ed. D.~J. {McLean} \& N.~R. {Labrum}
  (Cambridge and New York, Cambridge University Press), 289--332

\bibitem[{{Tasse} {et~al.}(2013){Tasse}, {van der Tol}, {van Zwieten}, {van
  Diepen}, \& {Bhatnagar}}]{2013A&A...553A.105T}
{Tasse}, C., {van der Tol}, S., {van Zwieten}, J., {van Diepen}, G., \&
  {Bhatnagar}, S. 2013, \aap, 553, A105, 1212.6178

\bibitem[{{The HDF Group}(1997-2014)}]{HDF5}
{The HDF Group}. 1997-2014, {Hierarchical Data Format, version 5},
  http://www.hdfgroup.org/HDF5/

\bibitem[{{van Haarlem} {et~al.}(2013){van Haarlem}, {Wise}, {Gunst}, {Heald},
  {McKean}, {Hessels}, {de Bruyn}, {Nijboer}, {Swinbank}, {Fallows},
  {Brentjens}, {Nelles}, {Beck}, {Falcke}, {Fender}, {H{\"o}randel},
  {Koopmans}, {Mann}, {Miley}, {R{\"o}ttgering}, {Stappers}, {Wijers},
  {Zaroubi}, {van den Akker}, {Alexov}, {Anderson}, {Anderson}, {van Ardenne},
  {Arts}, {Asgekar}, {Avruch}, {Batejat}, {B{\"a}hren}, {Bell}, {Bell}, {van
  Bemmel}, {Bennema}, {Bentum}, {Bernardi}, {Best}, {B{\^i}rzan}, {Bonafede},
  {Boonstra}, {Braun}, {Bregman}, {Breitling}, {van de Brink}, {Broderick},
  {Broekema}, {Brouw}, {Br{\"u}ggen}, {Butcher}, {van Cappellen}, {Ciardi},
  {Coenen}, {Conway}, {Coolen}, {Corstanje}, {Damstra}, {Davies}, {Deller},
  {Dettmar}, {van Diepen}, {Dijkstra}, {Donker}, {Doorduin}, {Dromer}, {Drost},
  {van Duin}, {Eisl{\"o}ffel}, {van Enst}, {Ferrari}, {Frieswijk}, {Gankema},
  {Garrett}, {de Gasperin}, {Gerbers}, {de Geus}, {Grie{\ss}meier}, {Grit},
  {Gruppen}, {Hamaker}, {Hassall}, {Hoeft}, {Holties}, {Horneffer}, {van der
  Horst}, {van Houwelingen}, {Huijgen}, {Iacobelli}, {Intema}, {Jackson},
  {Jelic}, {de Jong}, {Juette}, {Kant}, {Karastergiou}, {Koers}, {Kollen},
  {Kondratiev}, {Kooistra}, {Koopman}, {Koster}, {Kuniyoshi}, {Kramer},
  {Kuper}, {Lambropoulos}, {Law}, {van Leeuwen}, {Lemaitre}, {Loose}, {Maat},
  {Macario}, {Markoff}, {Masters}, {McFadden}, {McKay-Bukowski}, {Meijering},
  {Meulman}, {Mevius}, {Middelberg}, {Millenaar}, {Miller-Jones}, {Mohan},
  {Mol}, {Morawietz}, {Morganti}, {Mulcahy}, {Mulder}, {Munk}, {Nieuwenhuis},
  {van Nieuwpoort}, {Noordam}, {Norden}, {Noutsos}, {Offringa}, {Olofsson},
  {Omar}, {Orr{\'u}}, {Overeem}, {Paas}, {Pandey-Pommier}, {Pandey}, {Pizzo},
  {Polatidis}, {Rafferty}, {Rawlings}, {Reich}, {de Reijer}, {Reitsma},
  {Renting}, {Riemers}, {Rol}, {Romein}, {Roosjen}, {Ruiter}, {Scaife}, {van
  der Schaaf}, {Scheers}, {Schellart}, {Schoenmakers}, {Schoonderbeek},
  {Serylak}, {Shulevski}, {Sluman}, {Smirnov}, {Sobey}, {Spreeuw}, {Steinmetz},
  {Sterks}, {Stiepel}, {Stuurwold}, {Tagger}, {Tang}, {Tasse}, {Thomas},
  {Thoudam}, {Toribio}, {van der Tol}, {Usov}, {van Veelen}, {van der Veen},
  {ter Veen}, {Verbiest}, {Vermeulen}, {Vermaas}, {Vocks}, {Vogt}, {de Vos},
  {van der Wal}, {van Weeren}, {Weggemans}, {Weltevrede}, {White}, {Wijnholds},
  {Wilhelmsson}, {Wucknitz}, {Yatawatta}, {Zarka}, {Zensus}, \& {van
  Zwieten}}]{2013A&A...556A...2V}
{van Haarlem}, M.~P. {et~al.} 2013, \aap, 556, A2, 1305.3550

\bibitem[{{Warmuth} \& {Mann}(2005)}]{2005LNP...656...49W}
{Warmuth}, A., \& {Mann}, G. 2005, in Lecture Notes in Physics, Berlin Springer
  Verlag, Vol. 656, Lecture Notes in Physics, Berlin Springer Verlag, ed.
  K.~{Scherer}, H.~{Fichtner}, B.~{Heber}, \& U.~{Mall}, 49

\bibitem[{{Wild}(1950)}]{1950AuSRA...3..541W}
{Wild}, J.~P. 1950, Australian Journal of Scientific Research A Physical
  Sciences, 3, 541

\end{thebibliography}
\else
\printbibliography
\fi

\end{document}